\documentstyle[epsf]{l-aa}
\begin{document}

\thesaurus{06(08.05.3, 08.09.2 XX Pyx, 08.09.3, 08.15.1, 08.22.2)}

\title{Towards a seismic model of the $\delta$ Scuti star XX Pyxidis}

\author{A.~A.~Pamyatnykh\inst{1,2,3},
W.~A.~Dziembowski\inst{1,2},
G.~Handler\inst{2},
\and H.~Pikall\inst{2}}

\offprints{ W.~A.~Dziembowski}

\institute{N. Copernicus Astronomical Center, Polish
           Academy of Sciences, Bartycka~18, PL--00--716~Warszawa, Poland
\and
Institute of Astronomy, University of Vienna, T\"urkenschanzstra\ss e~17,
A--1180~Wien, Austria
\and
Institute of Astronomy, Russian Academy of Sciences, Pyatnitskaya~48,
           109017~Moscow, Russia}
\date{Received date / Accepted date}
\maketitle
\markboth{A.~A.~Pamyatnykh~et~al.:
          An asteroseismologial model of XX~Pyxidis}{}

%%%%%%%%%%%%%%%%%%%%%%%%%%%%%%%%%%%%%%%%%%%%%%%%%%%%%%%%%%
\begin{abstract}

Frequencies of 13 oscillation modes in the star XX~Pyxidis (CD--24~7599)
are accurately measured but for none
of the modes the spherical harmonic degree ($\ell$) is known.
We present results of an attempt to
construct the
model whose low--$\ell$ mode frequencies
reproduce possibly close the observations.
Models are constrained by the mean photometric and
spectroscopic data for the star.
However, the strongest constraint on the effective temperture
is from  the requirement that the modes excited in
the star fall into the range of the modes driven by the opacity mechanism.

Our models are built with the standard stellar evolution code allowing
no overshooting from the convective core. Effects of rotation are taken
into account both in stellar evolution and in linear nonadiabatic oscillation
calculations. Uniform rotation rate and conservation of the global
angular momentum during evolution are assumed.

We find several distinct mode identifications and associated stellar
models leading to frequency fits of similar quality.
Determination of the $\ell$ values for some of the modes could
remove the ambiguity.
None of the fits is satifactory: the mean departures exceed
the mean observational frequency error by at least one order
of magnitude. The fits could be improved by means of adjusting
model parameters that were kept fixed. However, such effort will be meaningful
only after improving accuracy in calculation of the
effects of rotation in oscillation frequencies.


\end{abstract}

%%%%%%%%%%%%%%%%%%%%%%%%%%%%%%%%%%%%%%%%%%%%%%%%%%%%%%%%%%%%%%%%%%%
\section{Introduction }

Long--time network observations of Delta Scuti stars resulted in
establishing truly multimodal character of oscillations
in a few objects of this type.
The four objects with the largest number of modes detected are
FG~Virginis with 24 modes (Breger et~al. 1997),
4~Canum Venaticorum with at least 17 modes (Breger 1997),
XX~Pyxidis with 13 modes (Handler et~al. 1997a)
and BH Piscium with 13 modes (Mantegazza et~al. 1996).
Though these numbers of
modes in comparison with the Sun or certain oscillating white dwarfs
may appear small, reproducing mode frequencies with model calculation
represents a challenge.

What makes our task more difficult than in the solar and white--dwarf cases,
is the fact that the observed spectra
do not reveal patterns enabling mode identification. Modes detected in
$\delta$ Scuti stars are of low radial order and their frequencies do
not obey simple asymptotic relations. Further, these stars are rather rapid
rotators and rotational splitting is not equidistant. Multicolor photometry
may be used to determine the spherical harmonic degree, $\ell$.
Spectroscopy allows, in addition, to determine
the azimuthal order, $m$. In practice, however, reliable $\ell$ and $m$
values are only seldom available. Thus, mode identification in $\delta$~Scuti
stars cannot be done independently of calculations involving
construction of equilibrium models and their oscillation properties.
The aim is to construct a stellar model with frequencies  of
low--degree modes which fit the observed ones. In
analogy to helioseismology we use the name {\it seismic model}.

The ultimate purpose of seismic model construction
is testing the ingredients of stellar
evolution theory: its basic assumptions and the input microscopic physics.
Stellar evolution theory has reached an advanced
level some 20 to 30 years ago. However, important questions concerning
convection and rotation remain unanswered. For instance,
is the mixing--length theory sufficient for describing convective flux
in subphotospheric layers?
If so, what is the best choice for the mixing length parameter,
$\alpha$, depending on stellar parameters?
What is the extent of element penetration, $d_{\rm over}$,  from the convective
to the radiative regions?
What is the law of the angular momentum evolution?
What is the role of rotationally induced instabilities in chemical element
mixing?

In recent years considerable progress has been made in treatment
of the equation of state and opacity for stellar interiors
(see Christensen--Dalsgaard \& D\"appen 1993 and references there,
Rogers et al. 1996, Rogers \& Iglesias 1994 and references there,
Iglesias \& Rogers 1996, Seaton 1996). The uncertainties, however, are
difficult to estimate.
It is therefore important to use possibly diversified sets of
observational data for testing. Observers provide us very accurate frequency
measurements but a major effort is needed to connect these quantities
to other stellar characteristics and to parameters used in model construction.

In the next section we will outline methodology
of constructing seismic models, which,
we believe, is applicable to all multimode $\delta$~Scuti stars
in the Main Sequence phase of evolution. We should stress that our work is not
the first effort aimed at reproducing observed frequencies in objects of
this type. Goupil et~al. (1993) modelled GX Pegasi for which
five frequencies have been measured.
Breger et~al. (1995) and Guzik and Bradley (1995) constructed
models of FG Virginis, which approximately reproduced observations. Our target
accuracy is much higher -- we want to approach the level of the observational
error.

The rest of this paper concerns XX~Pyxidis (CD--24~7599),
which seemed to be a good choice for the first attempt in accurate
reproducing observed spectrum. Amongst most multimode $\delta$~Scuti stars
it is the hottest
and least evolved. In such stars the predicted spectrum of unstable modes is
relatively simple. With the progress of stellar evolution the
instability extends to mode of mixed pressure/gravity (p/g) character and to
pure g--modes. Separations in frequency between modes of the same $\ell\!>\!0$
decrease, which complicates mode identifcation. In post Main Sequence stars
the density of predicted spectra is so high that the task
of mode identification
will remain impossible unless we discover the clue to mode selection.
Furthermore, the uncertainty in the description of the subphotosperic
convection is less severe in this relatively hot star.

After presenting data in Section 3, we discuss in Section 4
constraints on mean parameters of the star. The methodology of
the simultaneous search for mode identification and refinement of
model parameters is described in Section 5. In the same section
several alternative solutions are presented. In Section 6
we explain why these are unreliable and in the last Section we
discuss prospects for construction of a 
still improved seismic model of the star.

%%%%%%%%%%%%%%%%%%%%%%%%%%%%%%%%%%%%%%%%%%%%%%%%%%%%%%%%%%%%%%%%%%%%%%%%
\section{Construction of seismic stellar models}

%%%%%%%%%%%%%%%%%%%%%%%%%%%%%%%%%%%%%%%%%%%%%%%%%%%%%%%%%%
\subsection{Principles}

The number of modes detected in individual $\delta$~Scuti stars is far too
small to attempt to determine radial structure directly from measured
frequencies. Everything we may hope to
achieve is to use the observed frequencies to determine global parameters
characterizing the star and certain parameters of the theory.
The equations of problem are obtained by equating observed and calculated
frequencies,

\begin{equation}
$$f_{j,\rm obs}=f_{j, \rm cal}({\ell}_j,m_j,n_j,{\vec P}_S,{\vec P}_T),$$
\end{equation}

\noindent where $j$ labels measured frequencies,
${\ell}_j,m_j,n_j$ are numbers identifying the mode,
${\vec P}_S$ gives the set parameters characterizing the model, and
${\vec P}_T$ gives the set of free parameters of the theory.

A comment is needed about $n_j$.
Nonradial modes excited in  $\delta$~Scuti stars
could be  p--modes,
g--modes, and modes of mixed character.
In the present application,
continuity of $f$ as function of $\vec P$  for specified $({\ell},m,n)$ is
essential. The avoided crossing effect causes that the continuity
is guaranteed if we define the $n$ values at ZAMS where the
g-- and p--mode spectra are separated in frequency.
Following Unno et~al. (1989)
we will denote with $n<0$ and $n>0$ modes which in ZAMS models are
consecutive g-- and p--modes, respectively.
For $\ell>1$ there is the fundamental mode which is denoted with $n=0$. With
this choice, in evolved models some of modes denoted with $n>0$
will have predominatly g--mode properties and {\it vice versa}.

The parameters given by $\vec P_S$ are those characterizing the evolutionary
sequence like initial mass $M$, chemical composition, $X_0$ and $Z_0$,
and the angular momentum. In addition, there is a single parameter
identifying the model within the evolutionary sequence. The age
is always a good choice. In the present application we use
the initial equatorial velocity $V_{\rm rot,0}$
instead of the angular momentum and $\log T_{\rm eff}$ instead of the age.
The latter choice is acceptable because
XX~Pyx is certainly in the expansion phase of the Main Sequence
evolution and $T_{\rm eff}$ is a monotonic function of the age. In any
case there are five parameters in $\vec P_S$.

The quantities one may include in $\vec P_T$ are mixing--length parameter
$\alpha$, overshooting distance $d_{\rm over}$, as well as parameters
characterizing
angular momentum evolution and mass loss. None of these will be included
in the present work.
Constraining these parameters is certainly the most important application of
asteroseismology. However, as we will see later in this paper, we are not
yet at this stage.

%%%%%%%%%%%%%%%%%%%%%%%%%%%%%%%%%%%%%%%%%%%%%%%%%%%%%%%%%%
\subsection{Tools}

\subsubsection{Evolution code}

Stellar evolution code we use was developed in its original
version by B. Paczy\'nski, R. Sienkiewicz and M. Koz{\l}owski.
It is in fact a modern version of B. Paczy\'nski's code
(Paczy\'nski 1969, 1970),
which is now written in a modular form enabling an easy implementation of
microscopic physics data from various sources.
In the present application we use most recent versions of OPAL opacity
(Iglesias \& Rogers 1996) and equation of state (Rogers et~al. 1996) data.
The nuclear reaction rates are the same as used by Bahcall and Pinsonneault
(1995).

The effect of averaged centrifugal force is taken into account in the
equation for hydrostatic equilibrium.  We assume uniform rotation and
conservation of global angular momentum. These are the simplest
assumptions which we are prepared to abandon if this is required
for a successful seismic model construction.
Similarily, we
are prepared to introduce overshooting and/or rotationally induced
mixing of the chemical elements outside of the convective core,
but these effects are
ignored in the present work.
With these assumptions the input parameters for sequences are total
mass, $M$, initial values for hydrogen abundance, $X_0$,
metal abundance, $Z_0$,
and the equatorial velocity $V_{\rm rot,0}$. Initial heavy element mixture is
that of Grevesse \& Noels (1993).
The assumed parameters for
convection are $\alpha=1$, $d_{\rm over}=0$. The choice of $\alpha$
is unimportant in the present application.

For selected values of $T_{\rm eff}$ the
code returns all needed parameters of the model such as $\log g_{\rm eff}$,
$V_{\rm rot}$ or $\Omega$ and the matrix of the coefficients needed for
calculation of linear nonadiabatic oscillation properties.

\subsubsection {Nonadiabatic oscillation code}

The code we use is a modified version of the code developed long ago by one
of us (Dziembowski 1977). The important modification is taking into account
the effect of the averaged centrifugal force. Modified equations for adiabatic
oscillations are given by Soufi et~al. (1997). The corresponding change in
the full nonadiabatic set is trivial.
The input to this code specifies the range of frequencies and the
maximum value of $\ell$. The standard output are the following characteristics
of the $m=0$ modes: complex dimensionless frequencies
$\sigma$, modified growth rates, $\eta$, and the complex ratio of the flux
to displacement eigenfunctions at the surface, $\cal F$.

The frequency, $f$, which we compare with observations is given by the real
part of $\sigma$:

\begin{equation}
$$f=\Re(\sigma)\sqrt{G\bar\rho\over\pi}.$$
\end{equation}

For higher order p--modes beginning with, say,  $n=5$ or 6,
the nonadiabatic effects are important at the level of the
accuracy of frequency measurements. The growth rate is given by $-\Im(\sigma)$.
A more convenient measure of mode stability is $\eta$ which varies from $-\!1$,
if damping occurs \mbox{everywhere} in the stellar interior, to $+\!1$,
if driving occurs everywhere.
At neutral stability we have $\eta = 0$. Values of $\eta$ are
important as constraints on modes and models. We will use them in the
next sections.
Also $\cal F$'s have applications in asteroseismology, especially, in
determination of $\ell$--values (see Cugier et~al. 1994). However, we do not
have neccesary observational data on XX~Pyx to make use of this
quantity.

\subsubsection{Calculation  of the rotational splitting}

Rotation has an important effect on the structure of oscillation spectra.
Even at modest equatorial velocities such as 50--100 km/s the effect of
rotation can not be reduced to the linear splitting $\propto\!m\Omega$.
Our code calculating the rotational splitting is accurate
up to $\Omega^2$.
It is a version of the code by Dziembowski and Goode (1992) modified
by two of us (WAD, AAP) and M.--J.~Goupil.
It uses the adiabatic approximation.
The same nonadiabatic correction evaluated with the
the code described in the previous subsection is added to all modes within
the multiplet.

%%%%%%%%%%%%%%%%%%%%%%%%%%%%%%%%%%%%%%%%%%%%%%%%%%%%%%%%%%%%%%%%%%%
\section{Frequency data}

More than 350 hours of time--series photometric observations of XX~Pyx are
available. These include two Whole Earth Telescope runs (Handler et
al.~1996, 1997a) as well as follow--up measurements to study the star's
amplitude and frequency variations
(Handler et~al.~1997b).

Analysis of the two WET data sets allowed the extraction of 13 pulsation
frequencies plus one 2f--harmonic. These frequencies are summarized in
Table~\ref{freqobs},
and mostly taken from Handler et~al.~(1997a, hereafter HPO). Moreover, from an
examination of the follow--up data, the values of frequencies $f_1$, $f_2$ and
$f_3$ could be refined. The three revised frequencies are also listed in
Table~\ref{freqobs}.
Their error sizes differ from those listed in HPO, since these
frequencies are now known without alias ambiguity. However, these three
frequencies are slightly variable (Handler et~al.~1997b). Therefore the error
bars have been modified to account for the latter effect.

\begin{table}
   \caption[]{Frequencies of the 13 pulsation modes and the $2f_{1}$--harmonic
   of the $\delta$~Scuti star XX~Pyx unambiguously detected by HPO}
   \label{freqobs}
   \begin{tabular}{lccc}
   \hline
    & Frequency & Ampl.~(1992) & Ampl.~(1994)\\
    & (cycles/day) & (mmag) & (mmag)\\
   \hline
    $f_{1}$ & 38.1101 $\pm$ 0.0004 & 11.5 $\pm$ 0.2 & 15.9 $\pm$ 0.2\\
    $f_{2}$ & 36.0113 $\pm$ 0.0010 & 10.0 $\pm$ 0.2 & 3.3 $\pm$ 0.2\\
    $f_{3}$ & 33.4370 $\pm$ 0.0002 & 6.4 $\pm$ 0.2 & 3.8 $\pm$ 0.2\\
    $f_{4}$ & 31.3925 $\pm$ 0.0025 & 3.8 $\pm$ 0.2 & 1.9 $\pm$ 0.2\\
    $f_{5}$ & 28.9950 $\pm$ 0.0025 & 2.4 $\pm$ 0.2 & 2.7 $\pm$ 0.2\\
    $f_{6}$ & 27.0028 $\pm$ 0.0025 & 2.4 $\pm$ 0.2 & 1.6 $\pm$ 0.2\\
    $f_{7}$ & 34.6657 $\pm$ 0.0038 & 2.2 $\pm$ 0.2 & 2.0 $\pm$ 0.2\\
    $f_{8}$=$2f_{1}$ & 76.2203 $\pm$ 0.0038 & 0.8 $\pm$ 0.2 & 1.1 $\pm$ 0.2\\
    $f_{9}$ & 27.2635 $\pm$ 0.0138 & 1.0 $\pm$ 0.2 & 1.9 $\pm$ 0.2\\
    $f_{10}$ & 29.6196 $\pm$ 0.0075 & 1.5 $\pm$ 0.2 & 1.3 $\pm$ 0.2\\
    $f_{11}$ & 33.6367 $\pm$ 0.0125 & 1.2 $\pm$ 0.2 & 1.7 $\pm$ 0.2\\
    $f_{12}$ & 28.6980 $\pm$ 0.0138 & 1.1 $\pm$ 0.2 & 1.5 $\pm$ 0.2\\
    $f_{13}$ & 31.2033 $\pm$ 0.0163 & 1.4 $\pm$ 0.2 & 1.1 $\pm$ 0.2\\
    $f_{14}$ & 31.9062 $\pm$ 0.0100 & 1.0 $\pm$ 0.2 & 1.7 $\pm$ 0.2\\
   \hline
   \end{tabular}
\end{table}

%%%%%%%%%%%%%%%%%%%%%%%%%%%%%%%%%%%%%%%%%%%%%%%%%%%%%%%%%%%%%%%%%%%
\section{Constraints}
%
%%%%%%%%%%%%%%%%%%%%%%%%%%%%%%%%%%%%%%%%%%%%%%%%%%%%%%%%%%
\subsection{Mean surface parameters from observations}
To estimate the position of XX~Pyx in the H--R diagram, color photometry and
high--resolution spectroscopy of the star have been carried out by HPO.
Applying
calibrations to the photometric data, they infer that XX~Pyx is a
main--sequence $\delta$~Scuti star with $T_{\rm eff}=8300\pm200$~K, log
$g=4.25\pm0.15$. From their spectroscopy HPO determined the star's projected
rotational velocity with $v \sin i=52\pm2$ km/s and that the object has
approximately solar metal abundance: $[M/H]=0.0\pm0.2$.

%
%%%%%%%%%%%%%%%%%%%%%%%%%%%%%%%%%%%%%%%%%%%%%%%%%%%%%%%%%%
\subsection{Structures in the power spectra}

An important constraint for identifying the different pulsation frequencies
of XX~Pyx with the corresponding quantum numbers was found by HPO.
They searched for characteristic spacings within the
13 pulsation frequencies with a Fourier technique: they assumed unit amplitude
for all the individual pulsations they detected and Fourier analysed the
resulting signal. This is analogous to a spectral window: the power spectrum of
such a signal has maxima at the sampling frequency and its harmonics (see Kurtz
1983). With this method it is possible to find regular frequency spacings in
an objective way.

There are two physical reasons why regular frequency spacings can be present:
in case of slow (rigid) rotation the members of different multiplets are
approximately equally spaced in frequency. Since the measured $v \sin i$ of
XX~Pyx is 52 km/s (HPO), rotational splitting could, under favourable
circumstances, be detected. The second possibility is that consecutive modes
of the same $\ell$ are excited. Even for
low--overtone pulsations such modes are approximately equally spaced in
frequency.

\begin{figure}
  \epsfverbosetrue
  \begin{center}
  \epsfxsize=88mm
  \epsfysize=115mm
  \leavevmode
  \epsffile{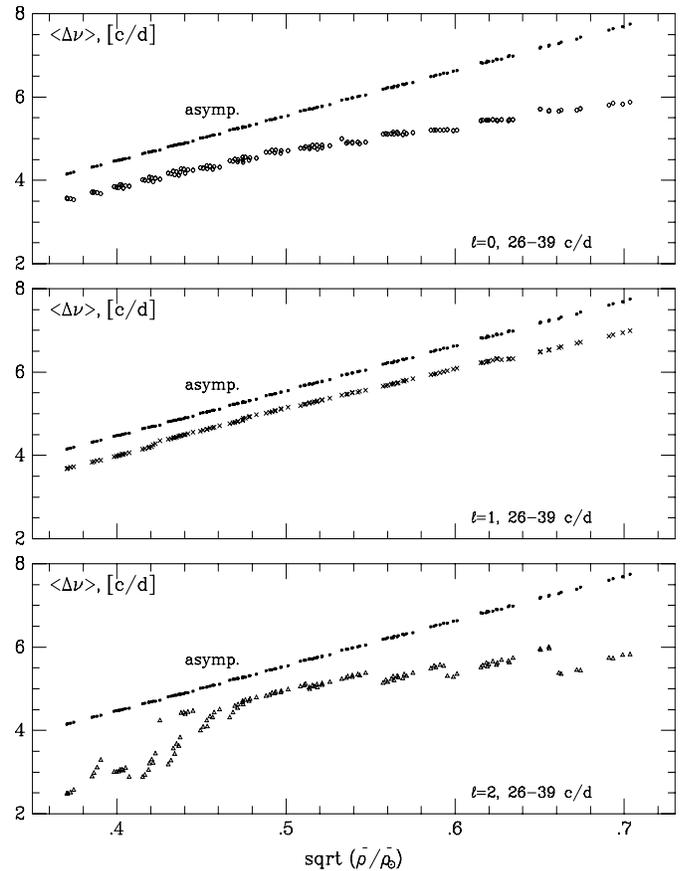}
  \end{center}
  \caption[]{Asymptotic frequency spacing of high-order p--modes of our
  main--sequence models compared to
  the mean spacing of \mbox{$\ell$ = 0, 1, 2} modes
  in the frequency range excited in XX~Pyx.
  Only axisymmetric ($m=0$) modes are shown.
  Rightmost points correspond to stellar models of lowest mass and
  highest effective temperature.
  The higher mass and/or the lower effective temperature the higher overtones
  of p--modes are appeared in a fixed frequency range, therefore the mean
  frequency spacing is approaching the asymptotic value.
  The scatter of the datapoints, most pronounced for $\ell = 0$ modes,
  reflects the different rotation rates of our models.
  The discontinuities of $\ell = 2$ modes
  at $\sqrt{(\bar\rho / \bar\rho_{\odot})} \leq 0.46$ 
  are due to intruding g--modes,
  while more regular level shifts of $\ell = 1$ and $\ell = 2$ modes
  are caused by different overtones (and different number of these overtones)
  appearing in the frequency range of interest.}
  \label{fig01}
\end{figure}

To illustrate this second possibility, we used our basic grid of
the main--sequence models and their oscillation frequencies
(see subsection 5.2):
$M = 1.75 - 2.05\ M_{\odot}$ with a step of $0.05\ M_{\odot}$,
$\log T_{\rm eff} = 3.905 - 3.925$ with a step of 0.05,
$V_{\rm rot,0} = 50 - 125$ km/s with a step of 25 km/s.
For each of these models, we compared the asymptotic frequency
spacing with the mean spacing between frequencies of
consecutive modes of a given $\ell$ value. We restrict to the observed
frequency range of XX~Pyx. The results are shown in Fig.~\ref{fig01}.
(Remember that the asymptotic frequency spacing for high overtone p--modes
is determined by the following formula, see Unno et al. 1989, Eq. (16.36):
$\Delta\nu=0.5/(\int_0^R {{dr} \over {c}})$, where $c$ is the sound speed,
--- i.e. it is proportional to the square root of the star's mean density.)

% ******************** Figure 1: xxfig01.ps

As can easily be seen from Fig.~\ref{delnu},
the assumption that regular frequency
spacings are present even at relatively low overtones is justified.
The frequency spacings of the different $\ell$'s are very similar.
Consequently, HPO carried out the Fourier analysis described above and
discovered a regular frequency spacing within the pulsation
modes of XX~Pyx, with a probability of less than 3\% that this detection
was caused by chance. Because of the value of this spacing ($\approx$ 26
$\mu$Hz\ $\approx2.25$ c/d), HPO suggested
that it cannot reflect rotational splitting (since the second--order terms
would destroy any frequency symmetry at such high rotation rates).

From a comparison of the Fourier analyses of the observed and of
several model frequency spectra, HPO concluded that this
spacing is caused by the presence of modes corresponding to
consecutive overtones of $\ell=1$ and $\ell=2$ modes.
HPO determined the spacing of consecutive
overtones of the {\it same} $\ell$ with
$54.0\pm2.3~\mu$Hz\ ($=4.67\pm0.20$~c/d). This frequency
spacing is a measure of the sound crossing time through the object, and thus
it can be used to determine the mean density of the star. For a model sequence
with solar metallicity, HPO determined
$\bar\rho = 0.246 \pm 0.020~\bar\rho_{\odot}$.

HPO performed several tests to verify the validity of this method and the
correct interpretation of its result (see their paper for more details) and
found their Fourier technique to be reliable, since the frequency spectrum of
XX~Pyx is very well suited for such an analysis: only a few
overtones are excited, but a large number of possible pulsation modes are
photometrically observed.

Actually, one can use this Fourier method to test a common assumption in mode
identification attempts from photometric (and radial velocity) data: with
increasing degree $\ell$ of the pulsations, geometrical cancellation
(Dziembowski 1977, Goupil et~al.~1996) decreases the photometric amplitude of
the modes. Therefore, it is usually assumed that only modes with $\ell<3$ can
be detected using photometric data. The photometric amplitude of an
$\ell=3$ mode with the same intrinsic amplitude as an $\ell=1$ mode is only
about a factor of 12 smaller, which is approximately the range of photometric
amplitudes of the different pulsation modes detected for XX~Pyx. However,
when considering rotational splitting, seven modes of $\ell=3$ can be present,
but only three $\ell=1$ modes. Therefore it can not be ruled out that a
non--negligible number of $\ell=3$ modes is excited to observable amplitude.

Consequently, we applied HPO's Fourier technique to model frequencies,
incorporating modes with $\ell=3$ and/or $\ell=4$. In cases where many of these
high $\ell$ modes were present, no significant frequency spacing could be
found, since their rotationally split patterns ``masked'' the regular spacing
of the $\ell=0-2$ modes. Only with a few of the high $\ell$ modes the typical
frequency spacing could be revealed. This suggests that most of the pulsation
modes of XX~Pyx are indeed of $\ell=0-2$.

%
%%%%%%%%%%%%%%%%%%%%%%%%%%%%%%%%%%%%%%%%%%%%%%%%%%%%%%%%%%
\subsection{Implication from the stability survey}
%

% ******************** Figure 2: xxfig02.ps
\begin{figure}
  \epsfverbosetrue
  \begin{center}
  \epsfxsize=88mm
  \epsfysize=100mm
  \leavevmode
  \epsffile{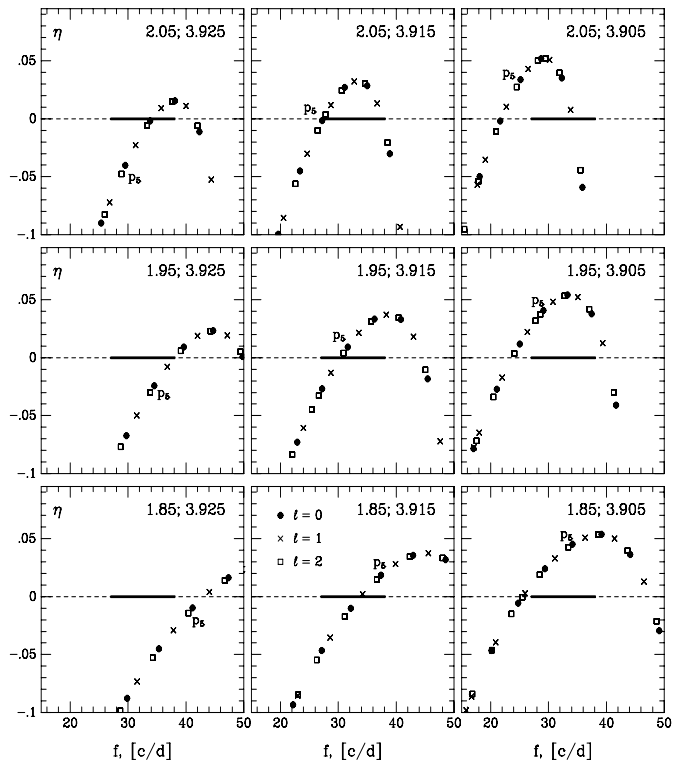}
  \end{center}
  \caption[]{Normalized growth rates, $\eta$, plotted against frequency, $f$,
  in selected models spanning the allowed ranges of $M$ and $\log T_{\rm eff}$.
  The values of $M$ (in solar units) and $\log T_{\rm eff}$ are given in each
  panel.
  Models were calculated with $X_0=0.7$, $Z_0=0.02$ and $V_{\rm rot}=0$.
  The symbol $\rm p_5$ is plotted near the corresponding radial overtone.
  The thick horizontal line shows the range of
  frequencies measured in XX~Pyx.}
  \label{fig02}
\end{figure}

XX~Pyx is a hot $\delta$~Scuti star located near the blue edge of
the instability strip. In such an object the opacity mechanism driving is
confined to modes in a narrow range of frequencies and the driving effect is
marginal. This is seen in Figure \ref{fig02}.
Values of $\eta$ never exceed 0.05.
The range of $f$ where $\eta>0$ never spans more than three
radial orders for radial modes. In models considered the lowest
values of $n$ is 4 and the highest is 8.
The behaviour of $\eta(f)$ in several models is shown in Fig.~\ref{fig02}.
The selected models have the surface parameters in the range consistent with
the data on XX~Pyx.
The independence on $\ell$ and occurence of a single maximum are typical
features for the opacity driven modes.

Values of $\eta$ do not have the same reliability as those of $f$.
Our treatment of the optically thin layers and the convective flux
is crude.
Therefore we will regard identifications with modes of small
negative $\eta$'s as admissible. However, we can exclude identifications
involving radial modes with $n<4$ and $n>8$
as well we can exclude all models
with, approximately, $\log T_{\rm eff}>3.925$ and the low mass models with
$\log T_{\rm eff}> 3.92$ (see Fig.~\ref{fig02}).

%%%%%%%%%%%%%%%%%%%%%%%%%%%%%%%%%%%%%%%%%%%%%%%%%%%%%%%%%%%%%%%%%%%
\section{A search for best parameters and mode identification}
We have no observational information about the $\ell$ and $m$ values for the
modes detected in XX~Pyx. The assignment of these two quantum numbers as
well as $n$ may only be done together with model parameter determination.
The basis is Eq. (1) applied to the 13 measured
frequencies. We only assume that all modes are of low degree. Most of
the results we will present here were obtained with
the assumption $\ell \le 2$ for all modes. We carried out also calculation
allowing one of the  mode to have $\ell=3$. Some of the results for such
a case will be discussed at the end of this section.

At this stage we allow no free theoretical parameters ${\vec P}_T$
and we fix $X_0$ and $Z_0$ at the values 0.7 and 0.02, respectively.
Thus, we allow only three free model parametres:
$M$, $V_{\rm rot,0}$, and $T_{\rm eff}$. For each model the modes
are assigned to minimize

\begin{equation}
$$\chi^2 ={1\over j}\sum_{i=1}^{j} (f_{{\rm obs},i}-f_{{\rm calc},i})^2,$$
\end{equation}

\noindent where $j=13$ or 12.
The first value was used in the case when we assumed
$\ell \le 2$ for all modes. In the alternative version we search for
the minimum of $\chi^2$ relaxing the fit for one of measured frequency.
In the automatic fitting
routine we made sure that each of the model frequencies was assigned at most
to one of the stellar frequencies.

%%%%%%%%%%%%%%%%%%%%%%%%%%%%%%%%%%%%%%%%%%%%%%%%%%%%%%%%%%
\subsection{Uncertainties in stellar and model frequencies}
In our search for best models and mode identification we
use unweighted observational data. The reason is that the
errors in the frequency measurements are smaller than the
uncertainities in the theoretical values.
In Table~\ref{freqobs} we may see that the errors range from $4\times 10^{-4}$
for $f_1$ to $163\times 10^{-4}$ for $f_{13}$. Thus the relative accuracy
of frequency is, at worst, $\sim 5\times 10^{-4}$, which is by one order
of magnitude less than the theoretical uncertainty.

The main problem on
the side of theory is in the treatment of rotation. We found that
at equatorial velocities $\sim 100$ km/s the second
order perturbation treatment yields relative accuracy in
frequencies $\sim5\times 10^{-3}$.

%%%%%%%%%%%%%%%%%%%%%%%%%%%%%%%%%%%%%%%%%%%%%%%%%%%%%%%%%%
\subsection{Tabulation of model frequencies}

With the codes described in subsection 2.2 we prepared tables
with frequencies for modes with $\ell=0$, 1, 2 covering the range of
the stellar pulsation frequencies. At this stage we neglected the effects of
near--degeneracy which we discuss in Sect.~6,
because it complicates calculations and it is not essential at this
stage.
The range of $\log T_{\rm eff}$ values was $[3.905 - 3.925]$ with a
step of 0.05. The lower limit is somewhat less than the lower limit allowed
by the photometry (subsection 4.1) while the upper limit follows from
the stability consideration (subsection 4.3).
The range of equatorial velocities was $[50 - 125]$ km/s with a step of
25 km/s. Here the lower limit follows from the $v\sin i$ measurement
(subsection 4.1). The upper limit was adopted
to avoid large errors in treatment of rotation.
The adopted mass range was $[1.75 - 2.05]M_{\odot}$ with a step of
$0.05M_{\odot}$.
The implied range of the mean density is significantly larger than that
inferred in subsection 4.2.
We consider models in the wider range of mean densities because there
is a nonzero probability that the HPO spacing is caused by chance.

%%%%%%%%%%%%%%%%%%%%%%%%%%%%%%%%%%%%%%%%%%%%%%%%%%%%%%%%%%
\subsection{Properties of $\chi^2(M,V_{\rm rot,0},T_{\rm eff})$}

% ******************** Figure 3: xxfig03.ps
\begin{figure}
  \epsfverbosetrue
  \begin{center}
  \epsfxsize=88mm
  \epsfysize=75mm
  \leavevmode
  \epsffile{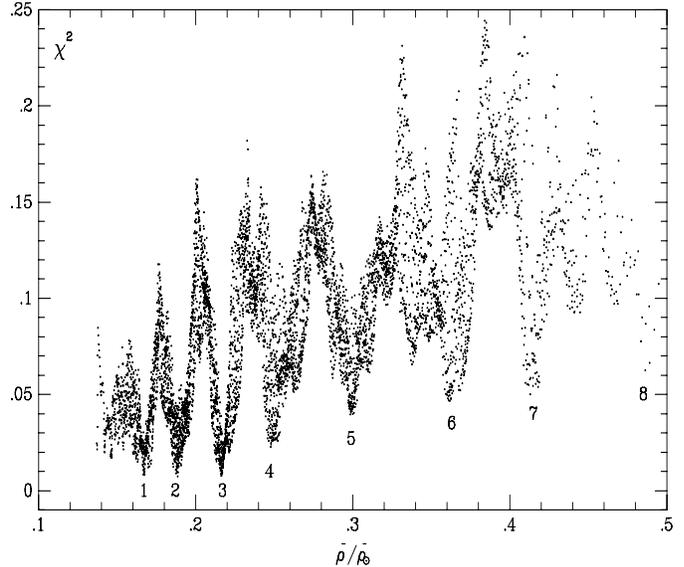}
  \end{center}
  \caption[]{Values of $\chi^2$ evaluated for frequencies in $\sim 7000$
              models in the ranges
              $M=1.75-2.05~M_{\odot}$,
              $V_{\rm rot,0}=50-110$ km/s,
              $\log T_{\rm eff}=3.905-3.925$.
              The patterns are the same (but less dense) as in the case
              of $\sim 40000$ models used to choose best identifications.}
  \label{fig03}
\end{figure}

\noindent We calculated values of $\chi^2$ according to Eq.(3) with $j=13$ for
nearly 40000 models interpolating model parameters and frequencies
from the three--dimensional basic grid described above.
At this stage we confined the upper
limit of the equatorial rotational velocity still more, to the value of
110 km/s.
In Fig.~\ref{fig03} we plot the values of
$\chi^2$ against mean density, which is the most important parameter
determining p--mode spectra. We see several dips in $\chi^2$ which
allows us to isolate
potential mode identification.
One of the dips occurs close to $\bar\rho = 0.246$, which was
the preferred value from the observational frequency spacing. The present
analysis shows that there are few alternative "good" values.
We stress that a good fit may occur even if there is no preferred
spacing. The coincidence however speaks in favor of identfications
corresponding to this particular dip.
The dips
become shallower with the increasing $\bar\rho$ which merely reflects
the fact that in less evolved models there are fewer modes in the
specified frequency range.

At this stage we have to consider
the identifications associated with the dips
located near the following eight values of $\bar\rho$: 0.17, 0.19, 0.22, 0.25,
0.30, 0.36, 0.41 and 0.49.
It should be noted, however, that all the minimum values
of $\chi^2$ are still far larger than the measurement errors.

Next important parameter is the equatorial velocity of rotation
which determines the
$m$--dependence. The plot of $\chi^2$ against $V_{{\rm rot},0}$, which
we do not reproduce here, shows two shallow minima around
the values 60 and 100 km/s.

%%%%%%%%%%%%%%%%%%%%%%%%%%%%%%%%%%%%%%%%%%%%%%%%%%%%%%%%%%
\subsection{Possible models and identifications}

There is no doubt that if all (or almost all) 13 modes detected in
XX~Pyx are of $\ell \le 2$ degree then the mode identification
and model must correspond to one of the dips in $\chi^2$.
At each dip we chose a number of identifications with nearly the same lowest
values of $\chi^2$. All such identifications are listed in Table~\ref{ident}.
They are grouped according to the dips. For each specified identification
the three model parameters -- $M$, $V_{\rm rot,0}$, $\log T_{\rm eff}$ --
are determined through the $\chi^2$ minimalization. These and other
model characteristics are listed in Table~\ref{models}.

\begin{table*}
   \begin{center}
   \caption[]{Possible identifications of the 13 modes detected in
    XX~Pyx. Horizontal lines separate groups of the identifications
    associated
    with the consecutive dips of $\chi^2$ seen in Fig.~\ref{fig03}.
    Note that within
    the group most of the identifications are identical.
    Models corresponding to each identification are determined 
    by minimalization of $\chi^2$.}
   \label{ident}
   \begin{tabular}{lccccccccccccc}
\hline

    & $f_{6}$ & $f_{9}$ & $f_{12}$ & $f_{5}$ & $f_{10}$ & $f_{13}$ & $f_{4}$ &
$f_{14}$ & $f_{3}$ & $f_{11}$ & $f_{7}$ & $f_{2}$ & $f_{1}$\\

       & 27.003& 27.264& 28.698& 28.995& 29.620& 31.203& 31.392& 31.906& 33.435& 33.637& 34.666& 36.012& 38.110\\

%Error &~~.002 &~~.014 &~~.014 &~~.002 &~~.008 &~~.016 &~~.002 &~~.010 &~~.002 &~~.012 &~~.004 &~~.002 &~~.001 \\

Model & l n m & l n m & l n m & l n m & l n m & l n m & l n m & l n m & l n m & l n m & l n m & l n m & l n m \\
\hline
  D1A & 2 3-1 & 2 3-2 & 1 5 1 & 2 4-1 & 1 5-1 & 2 5 0 & 2 5-1 & 2 5-2 & 1 6 0 & 1 6-1 & 2 6 1 & 2 6-2 & 2 7 2 \\
  D1B & 2 3-1 & 2 3-2 & 1 5 1 & 2 4-2 & 1 5-1 & 2 5 0 & 0 6 0 & 2 5-2 & 1 6 0 & 1 6-1 & 2 6 1 & 2 6-2 & 2 7 2 \\
  D1C & 2 3-1 & 2 4 2 & 1 5 1 & 2 4-2 & 1 5-1 & 2 5 0 & 2 5-1 & 2 5-2 & 1 6 0 & 1 6-1 & 2 6 1 & 2 6-2 & 2 7 2 \\
  D1D & 2 3-2 & 2 4 2 & 1 5 1 & 2 4-2 & 1 5-1 & 2 5 0 & 2 5-1 & 2 5-2 & 1 6 0 & 1 6-1 & 2 6 1 & 2 6-2 & 2 7 2 \\
  D1E & 2 3-1 & 2 4 2 & 1 5 1 & 2 4-2 & 1 5-1 & 2 5 0 & 0 6 0 & 2 5-2 & 1 6 0 & 1 6-1 & 2 6 1 & 2 6-2 & 2 7 2 \\
  D1F & 2 3-2 & 2 4 2 & 1 5 1 & 2 4-2 & 1 5-1 & 2 5 0 & 0 6 0 & 2 5-2 & 1 6 0 & 1 6-1 & 2 6 1 & 2 6-2 & 2 7 2 \\
\hline
  D2A & 1 4-1 & 2 3-1 & 2 4 0 & 0 5 0 & 1 5 1 & 1 5 0 & 1 5-1 & 2 5 1 & 2 5-1 & 1 6 1 & 2 6 2 & 2 6 1 & 2 6-2 \\
  D2B & 1 4-1 & 2 3-2 & 2 4 0 & 0 5 0 & 2 4-2 & 1 5 0 & 1 5-1 & 2 5 1 & 2 5-1 & 1 6 1 & 2 6 2 & 2 6 1 & 2 6-2 \\
  D2C & 1 4-1 & 2 3-2 & 2 4 0 & 2 4-1 & 2 4-2 & 1 5 0 & 1 5-1 & 2 5 2 & 0 6 0 & 2 5-1 & 1 6 1 & 2 6 2 & 2 6-2 \\
\hline
  D3A & 2 3 0 & 1 4 1 & 2 4 2 & 1 4-1 & 2 4 1 & 0 5 0 & 2 4-1 & 1 5 1 & 1 5 0 & 1 5-1 & 2 5 1 & 2 5-1 & 1 6 0 \\
  D3B & 2 3-1 & 1 4 1 & 1 4 0 & 1 4-1 & 2 4 1 & 0 5 0 & 2 4-1 & 1 5 1 & 1 5 0 & 1 5-1 & 2 5 1 & 2 5-1 & 1 6 0 \\
  D3C & 2 3-1 & 1 4 1 & 2 4 2 & 1 4-1 & 2 4 1 & 0 5 0 & 2 4-1 & 1 5 1 & 1 5 0 & 1 5-1 & 2 5 1 & 2 5-1 & 1 6 0 \\
\hline
  D4A & 2 3 2 & 2 3 1 & 0 4 0 & 2 3-2 & 1 4 1 & 1 4-1 & 2 4 2 & 2 4 1 & 0 5 0 & 2 4-2 & 1 5 1 & 1 5-1 & 0 6 0 \\
  D4B & 2 3 2 & 2 3 1 & 2 3-1 & 2 3-2 & 1 4 1 & 1 4-1 & 2 4 2 & 2 4 1 & 0 5 0 & 2 4-2 & 1 5 1 & 1 5-1 & 0 6 0 \\
\hline
  D5A & 1 3 1 & 2 2-2 & 1 3-1 & 2 3 2 & 2 3 1 & 2 3-1 & 0 4 0 & 2 3-2 & 1 4 0 & 2 4 2 & 2 4 1 & 2 4 0 & 1 5 1 \\
\hline
  D6A & 2 2 2 & 2 2 1 & 2 2-1 & 2 2-2 & 1 3 1 & 1 3 0 & 1 3-1 & 2 3 2 & 2 3 0 & 2 3-1 & 2 3-2 & 1 4 1 & 2 4 2 \\
\hline
  D7A & 2 1-2 & 1 2-1 & 2 2 2 & 2 2 1 & 2 2 0 & 0 3 0 & 2 2-2 & 1 3 1 & 1 3 0 & 1 3-1 & 2 3 1 & 2 3 0 & 1 4 1 \\
  D7B & 2 1-2 & 1 2-1 & 2 2 2 & 2 2 1 & 2 2 0 & 0 3 0 & 2 2-2 & 1 3 1 & 1 3-1 & 2 3 2 & 2 3 1 & 2 3 0 & 1 4 1 \\
\hline
  D8A & 2 1 0 & 0 2 0 & 2 1-2 & 1 2 0 & 1 2-1 & 2 2 2 & 2 2 1 & 2 2 0 & 2 2-2 & 0 3 0 & 1 3 1 & 1 3 0 & 2 3 1 \\
\hline
   \end{tabular}
   \end{center}

\end{table*}

\begin{table*}
   \begin{center}
   \caption[]{Parameters of models determined by the $\chi^2$
     minimalization. $N_m$ is the total number of modes with $\ell\le2$ within
     the frequency range of modes detected in XX~Pyx. The number decreases
     with mean density which explains increase of the $\chi^2$ minima.
     Note that the difference in mean radial mode degree between consecutive
     groups is close to 0.5. Models used in Fig.~\ref{fig05} are marked
     with asterisks in the last column. }
   \label{models}
   \begin{tabular}{lcccccccccc}
\\
\hline
Model & $M/M_{\odot}$ & $V_{\rm rot}({\rm ZAMS})$ & $V_{\rm rot}$ & $\log T_{\rm eff}$
& $\log L$ & $\bar\rho/\bar\rho_{\odot}$ & $\chi^{2}$ & $N_m$ & $\Sigma (n+l/2)/13$\\
\hline
D1A & 1.9869 & 57.50 & 51.92 & 3.90533 & 1.2906 & .16692 & .006876 & 30 & 5.92\\
D1B & 2.0181 & 61.82 & 55.92 & 3.91101 & 1.3175 & .16717 & .006217 & 29 & 5.92\\
D1C & 2.0160 & 60.18 & 54.43 & 3.91063 & 1.3158 & .16710 & .006224 & 29 & 6.00\\
D1D & 2.0266 & 59.38 & 53.69 & 3.91247 & 1.3249 & .16708 & .007508 & 29 & 6.00\\
D1E & 2.0165 & 61.55 & 55.69 & 3.91072 & 1.3161 & .16715 & .006108 & 29 & 6.00 & $\ast$\\
D1F & 2.0260 & 60.59 & 54.79 & 3.91235 & 1.3243 & .16711 & .007646 & 29 & 6.00\\
\hline
D2A & 1.9953 & 108.01 & 99.30 & 3.91241 & 1.2876 & .18680 & .008070 & 27 & 5.73 & $\ast$\\
D2B & 2.0477 & 110.35 & 101.56 & 3.92169 & 1.3320 & .18681 & .007450 & 26 & 5.62\\
D2C & 2.0333 & 66.67 & 61.20 & 3.92056 & 1.3233 & .18842 & .007387 & 26 & 5.62\\
\hline
D3A & 1.9122 & 95.00 & 88.65 & 3.90567 & 1.2064 & .21608 & .007856 & 24 & 5.23 & $\ast$\\
D3B & 1.9986 & 97.64 & 91.16 & 3.92154 & 1.2819 & .21614 & .007181 & 23 & 5.23\\
D3C & 1.9972 & 95.40 & 89.09 & 3.92132 & 1.2809 & .21605 & .007530 & 23 & 5.23\\
\hline
D4A & 1.8722 & 66.67 & 63.00 & 3.90611 & 1.1635 & .24711 & .019895 & 20 & 4.77\\
D4B & 1.8782 & 67.27 & 63.60 & 3.90723 & 1.1688 & .24732 & .020005 & 20 & 4.77 & $\ast$\\
\hline
D5A & 1.8367 & 81.32 & 78.62 & 3.90907 & 1.1149 & .29872 & .036899 & 19 & 4.23 & $\ast$\\
\hline
D6A & 1.8527 & 60.28 & 59.07 & 3.92234 & 1.1148 & .36149 & .045757 & 18 & 3.69\\
\hline
D7A & 1.7556 & 75.00 & 74.63 & 3.90944 & 1.0074 & .41550 & .050614 & 17 & 3.27\\
D7B & 1.7562 & 74.13 & 73.72 & 3.90933 & 1.0086 & .41368 & .048300 & 17 & 3.31\\
\hline
D8A & 1.7500 & 61.11 & 61.17 & 3.91500 & 0.9819 & .48689 & .061828 & 18 & 2.85\\
\hline
   \end{tabular}
   \end{center}
\end{table*}

Not all models listed in Table~\ref{models} are consistent with
the requirement that the identified modes are pulsationally
unstable.
Thus far we made a limited use of the constraint following from the stability
considerations.

\begin{figure}
  \epsfverbosetrue
  \begin{center}
  \epsfxsize=85mm
  \epsfysize=110mm
  \leavevmode
  \epsffile{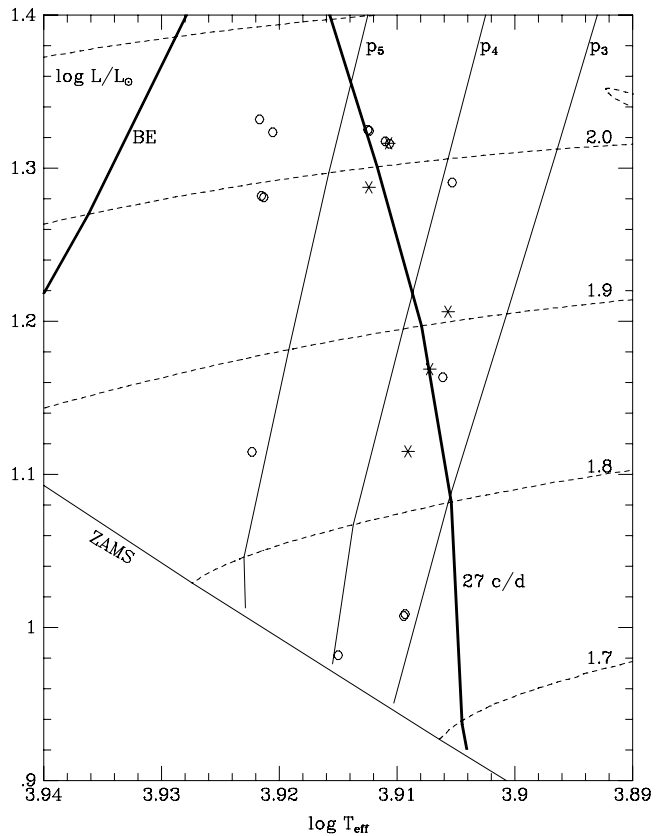}
  \end{center}
  \caption[]{Position (symbols) of the best models in the H--R diagram.
   The asterisks show the models used in Fig.~\ref{fig05}. The evolutionary
   tracks for models with indicated mass and $V_{\rm rot}=0$
   are shown with dashed lines.
   The line denoted with BE shows the absolute blue edge;
   there are no unstable modes in models to the left of this line.
   The lines denoted with p$_n$ are the blue edges for the corresponding
   radial overtones and the line denoted with 27 c/d is the blue edge for
   modes with the corresponding frequency.}
  \label{fig04}
\end{figure}

In fact, the stability argument may be used to
eliminate some of the models listed in Tables \ref{ident} and \ref{models}
as candidates
for the seismic model of XX Pyx. A comparison of the model positions
with the blue edges in the H--R diagram, shown in Fig.~\ref{fig04},
facilitates the elimination.
We see that all the models with $M\le 1.9M_\odot$ and  hotter than
some $\log T_{\rm eff}=3.91$ lie
rather far from the blue edge for modes with $f\le 27$ c/d, which
is the lowest frequency observed in XX Pyx. At higher masses the upper
limit of $\log T_{\rm eff}$ should be moved to 3.914. In this way we
may eliminate models D2B, D2C, D3B, D3C and D8A.

% ******************** Figure 4: xxfig04.ps

Perhaps model D1A may be
eliminated with the opposite argument because it lies far to the red of
the 27 c/d blue edge. This means that there are unstable modes in the
model with lower frequencies that are not seen in the star. This is,
however, a much weaker argument because it is not an uncommon situation
that unstable modes are not excited with detectable amplitudes.

% ******************** Figure 5: xxfig05.ps
\begin{figure}
  \epsfverbosetrue
  \begin{center}
  \epsfxsize=88mm
  \epsfysize=155mm
  \leavevmode
  \epsffile{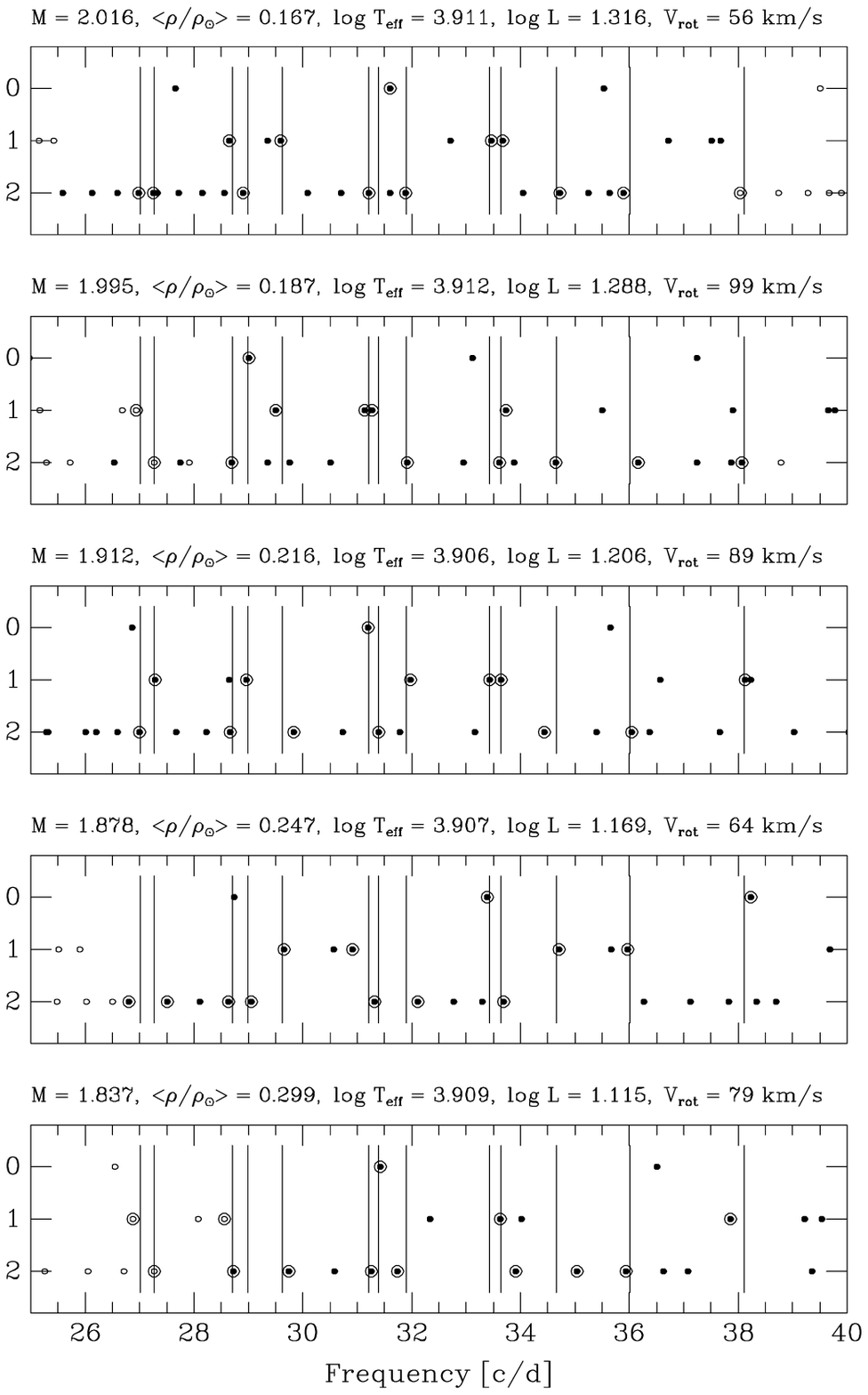}
  \end{center}
  \caption[]{Comparison of the XX Pyx frequencies (vertical lines)
  with model frequencies (symbols) for $\ell=0$ (top row),
  $\ell=1$ (middle row),  $\ell=2$ (bottom row). The encircled
  symbols denote the modes identified with those excited in
  the star. Small open circles denote the stable modes (always
  very close to instability $\eta\approx 0$. Positions of
  the selected models in the H--R diagram are denoted with asterisks in
  Fig.~\ref{fig04}. In Tables \ref{ident} and \ref{models} these models
  (from top to bottom) are denoted as
  D1E, D2A, D3A, D4B, and D5A.}
  \label{fig05}
\end{figure}

In Fig.~\ref{fig05} the frequencies determined for XX Pyx are compared with
frequencies of low degree modes in selected models. These are not
all acceptable but they cover the whole range. The common pattern of
the model frequency spectra is the large departure from the equidistant
pattern of the rotational splitting. This is particularly well seen
in the case of the $\ell=1$ triplets.
One may see that with the increasing
radial order the triplet becomes more and more asymmetric,
so that the prograde mode $m\!=\!-1$ almost overlap the centroid mode $m\!=\!0$
at the high frequency end.
There is a near equidistant
separation between the consecutive centroid modes at each $\ell$ in most
of the cases. The exception is the low frequency part of the $\ell=2$
spectrum in all, but the lowest mass models. Here we see the effect of
{\it avoided crossing} between two multiplets. Modes in this range
have mixed (p-- and g--) character. Their frequencies are sensitive to
the structure of the deep interior and therefore these modes are of special
interest for seismic sounding.

%%%%%%%%%%%%%%%%%%%%%%%%%%%%%%%%%%%%%%%%%%%%%%%%%%%%%%%%%%
\subsection{Fitting 12 modes}

Naturally, if we demand fitting for only 12 of the
measured frequencies we may attain significantly lower values of $\chi^2$.
To estimate the effect we looked for the minimum 
of $\chi^2$ considering all possible choices of 12 modes for each model.
As expected, the general patterns
of the $\chi^2$ dependence on $\bar\rho/\bar\rho_\odot$ 
were found very much like those
in Fig.~\ref{fig03}. In particular, the dips occured at the same locations.
For the same
set of models the absolute minimum of $\chi^2$ (0.00254) was reached at
dip~3 ($\bar\rho/\bar\rho_\odot=0.216$). The minimum value was nearly
three times smaller than the absolute minimum in the previous 13-mode case.
The implied mean  frequency mismatch of 0.05~c/d is close to the estimated
uncertainty in calculated frequencies. It is, thus, clear that improvement on
the side of theory is needed before we will be able to produce a
credible seismic model for this star.
In the next section we discuss two most needed improvements.

%%%%%%%%%%%%%%%%%%%%%%%%%%%%%%%%%%%%%%%%%%%%%%%%%%%%%%%%%%%%%%%%%%%
\section{Problems}
One important effect in the treatment of rotational frequency perturbation
which we did not include so far in our calculation was the
near--degeneracy of certain modes which may be coupled by rotation.
The effect is discussed in details by Soufi et~al. (1997). Here we
will present selected results for our model D3A.

Uniform (more generally, spherical) rotation couples modes if their
azimuthal numbers ($m$) are the same and their degrees ($\ell$) are
the same or differ by 2. If the frequency difference between the two (or more)
coupled modes
is of the order of the rotation frequency one has to use the version
of the peturbation formalism appropriate for the case of degeneracy
i.e. to consider as a zeroth order basis a linear combination of
the nearly degenerate modes. The relative contributions of the components
are determined from the second order (in $\Omega$) perturbation
equations.

In application to XX~Pyx a systematic near--de\-ge\-ne\-racy occurs between
the $\ell=0$ and 2 modes and between the $\ell=1$ and 3 modes. In the
latter case, for each multiplet we have three coupled modes corresponding
to $m=-1$, 0, 1. These near--degeneracies
follow from rather high values of $n$ and, hence, approximate
validity of the p--mode asymptotics.
We also considered coupling involving three modes, e.g. $\ell=0$, 2, 4
or $\ell=0$, 2, 2, in the case of avoided crossing. In none of the
cases considered the inclusion of the third mode was essential.

\begin{figure}
  \epsfverbosetrue
  \begin{center}
  \epsfxsize=88mm
  \epsfysize=95mm
  \leavevmode
  \epsffile{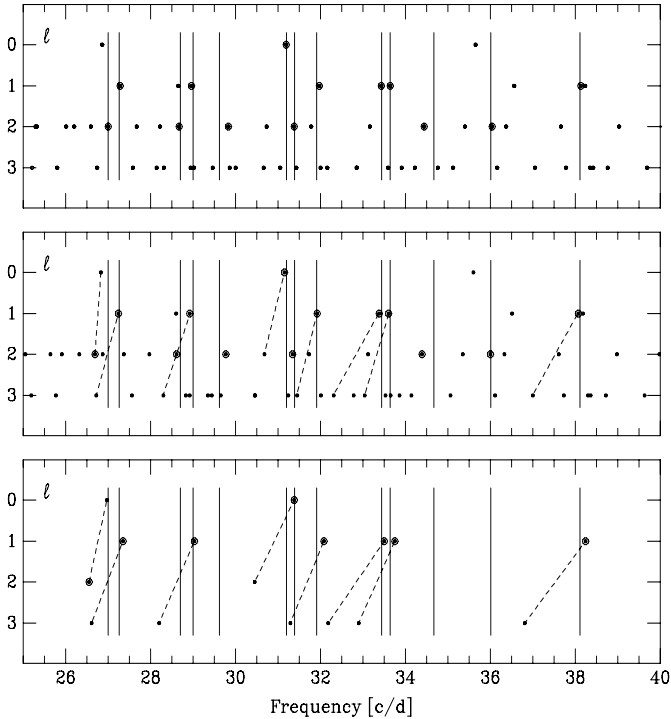}
  \end{center}
  \caption[ ]{Effects of inadequate grid density and near degeneracy
              shown for one of the best models (D3A). The upper panel shows
              interpolated frequencies which are the same (except that $\ell=3$
              modes are added) as in the mid panel in Fig.~\ref{fig05}.
              The mid panel here shows frequencies
              of the same modes for the recalculated model.
              Frequencies of the modes coupled by rotation
              are connected with dashed lines. The resulting frequencies of
              these modes are shown in the lower panel.}
  \label{fig06}
\end{figure}

Model D3A and its oscillation frequencies were obtained by means of
interpolation described in the previous section.
In order to evaluate the effect of the degeneracy we had
to recalculate the model and its frequencies.
A comparison of the two
upper panels of Fig.~\ref{fig06} clearly
shows that the grid of models
was not sufficiently dense. The differences between interpolated and
calculated frequencies are easily seen, especially
for the $\ell=2$ modes near the avoided crossing.
This inadequacy of our grid is not essential at this stage but must be
kept in mind in more advanced efforts. In the top panel one may see how
the fit is improved if the $\ell=2$ identification for $f_7=34.7$ c/d is
replaced with the $\ell=3$. The value of $\chi^2$ is then lowered by some
30 \%. No fit improvement is achieved by allowing the $\ell=3$ identification
for another poorly fit frequency $f_{10}=29.6$ c/d. One may see, however,
in the mid panel that the situation changes if instead of the interpolated
frequencies one uses the calculated ones.

A comparison of the mid and bottom panels shows the effect of the
coupling between nearly degenerate modes. The effect is best seen in
the case of $\ell=2$ and 0 modes. The frequency distance between the
modes increases when the coupling is taken into account.
In Table~\ref{coupling}
we give the values of the frequency shifts ($\delta f$) caused by the mode
coupling as well as the amplitudes of the spherical harmonic components
($A_{\rm low}$ is corresponding to $\ell=0$ or 1 and $A_{\rm high}$
is corresponding to $\ell=2$ or 3). One may see that the mutual
contamination of the $\ell=0$ and 2 components is quite strong.
This is bad news for prospects of mode discrimination by means
of two--color photometry because such contaminated modes may appear in
the {\it amplitude ratio -- phase difference} diagrams
(Watson 1988, Garrido et~al. 1990)
in the $\ell=0$, 1 or 2 domains
depending on the inclination of the rotation axis.
Fortunately, modes with $m\ne 0$ are not affected. The mutual contamination
is also significant for the $\ell=1$ and 3 pairs. In all the cases
the nominal $\ell=3$ modes will be most likely observable through their
$\ell=1$ contaminations.

% ******************** Figure 6: xxfig06.ps

\begin{table}
   \caption[]{Effects of coupling between nearly degenerate modes}
   \label{coupling}
   \begin{tabular}{lccccccc}
\hline
$f_{\rm obs}$& $\ell$ & $n$ & $m$ & $f_0$ & $\delta f$ & $A_{\rm low}$ & $A_{\rm high}$\\
\hline
$f_{6}$  & 0 & 4 &  0 & 26.832 &  0.139 & 0.923 & -0.385\\
         & 2 & 3 &  0 & 26.686 & -0.139 & 0.761 &  0.649\\
\hline
$f_{9}$  & 1 & 4 &  1 & 27.239 &  0.112 & 0.935 & -0.356\\
         & 3 & 3 &  1 & 26.720 & -0.112 & 0.422 &  0.906\\
\hline
$f_{5}$  & 1 & 4 & -1 & 28.932 &  0.096 & 0.951 & -0.308\\
         & 3 & 3 & -1 & 28.293 & -0.096 & 0.374 &  0.927\\
\hline
$f_{13}$ & 0 & 5 &  0 & 31.153 &  0.226 & 0.881 & -0.474\\
         & 2 & 4 &  0 & 30.678 & -0.226 & 0.513 &  0.858\\
\hline
$f_{14}$ & 1 & 5 &  1 & 31.923 &  0.153 & 0.915 & -0.403\\
         & 3 & 4 &  1 & 31.443 & -0.153 & 0.481 &  0.877\\
\hline
$f_{3}$  & 1 & 5 &  0 & 33.374 &  0.122 & 0.962 & -0.275\\
         & 3 & 4 &  0 & 32.302 & -0.122 & 0.337 &  0.942\\
\hline
$f_{11}$ & 1 & 5 & -1 & 33.613 &  0.137 & 0.931 & -0.365\\
         & 3 & 4 & -1 & 33.038 & -0.137 & 0.441 &  0.897\\
\hline
$f_{1}$  & 1 & 6 &  0 & 38.064 &  0.174 & 0.944 & -0.329\\
         & 3 & 5 &  0 & 36.988 & -0.174 & 0.370 &  0.929\\
\hline
   \end{tabular}
\end{table}

%%%%%%%%%%%%%%%%%%%%%%%%%%%%%%%%%%%%%%%%%%%%%%%%%%%%%%%%%%%%%%%%%%%%%%%%%%%
\section{Discussion}
Clearly, we have not succeeded in constructing the seismic model
of XX~Pyx. Models regarded as plausible, such as those in Fig.~\ref{fig05},
do not reproduce the frequencies within the observational errors.
In fact, we are quite far from the $10^{-3} - 10^{-2}$ c/d
error range seen in Table~\ref{freqobs}. The departures from the fit for
some of the identifications are in fact easily visible in Fig.~\ref{fig05}.
The problem proved more difficult than we have anticipated.
We believe that we learned something in the process and
that this knowledge should be shared with other groups undertaking
similar efforts.

Undoubtedly, an improvement of the fits could be accomplished
by allowing adjustment of the chemical composition
parameters. The overshooting distance and parameters in the
radial dependence of rotation rate should also be regarded as
adjustable quantities. These degrees of freedom affect primarily
the positions of the mixed modes relative to pure p--modes. In
this application the mixed modes exist for $\ell=2$ at the low
frequency end. Such modes extend to higher frequencies for
higher $\ell$'s. However, the freedom allows also a fine tuning of the
distances between pure p--modes. One could hope that there is
just one mode identification and one corresponding model of the star
for which the frequency fit within the measurement errors is possible.

The reason why we cannot yet  proceed this way is that our treatment
of rotation is not adequate. We used here a perturbation theory
which is accurate up to some 0.05 c/d, much worse than the measurement
accuracy. We are not yet sure that the recently developed cubic
theory (Soufi et~al. 1997) would yield a sufficient accuracy.

Observational determination of the $\ell$ values for some of
the excited modes could significantly change the situation. Imagine,
for instance, that we find that the photometric data place the $f_1$ mode
into the $\ell=0$ domain of an {\it amplitude ratio -- phase difference}
diagram (see Watson 1988, Garrido et~al. 1990).
We may see in Table~\ref{freqobs} that
this mode has far the highest amplitude and certainly it will be the first for
which we will have the required data.
As we have discussed in Section 6,
the possibility that this is not a $\ell=0$ mode but
rather a $\ell=2, m=0$ mode must be considered.
Fortunately, we do not see such an identification
for $f_1$ in Table \ref{ident}.
Thus, we are left
with only two models, D4A and D4B, with very similar parameters and the same
identification for almost all modes excited in XX~Pyx.
We could discriminate between the two if we knew $\ell$ for
$f_{12}$ which is the only mode with different identification.
If, however, the $f_1$ position in the diagram corresponds
to an $\ell=1$ domain, we are less lucky. The possible models are
not only six models with $\ell=1$ for $f_1$ but also models D4A and D4B
with $\ell=0$
because radial modes contaminated with a $\ell=2$ component may
appear in the $\ell=1$ domain. The ambiguity
implies a wide range of admissible model parameters.

%%%%%%%%%%%%%%%%%
\begin{acknowledgements}
Pamyatnykh and Dziembowski enjoyed
the hospitality during their stays in Vienna and express their thanks to
Mike Breger and his collaborators for it.
We all thank Bohdan Paczy\'nski, Maciej Koz{\l}owski and Ryszard Sienkiewicz
for the stellar evolution code. We thank also Marie--Jo Goupil for
her contribution to development of the code calculating the rotational
splitting. This work was supported in part by the
Austrian Fonds zur F\"{o}rderung der wissenschaftlichen Forschung,
project number S7304, and also by the grants AFOSR--95--0070, 
NSF--INT--93--14820, RFBR--95--02--06359 and KBN--2P304--013--07.

\end{acknowledgements}

\end{document}